\documentclass[aps,twocolumn,floats,prd,nofootinbib,showpacs,showkeys,letterpaper]{revtex4} 
\usepackage[dvips]{graphicx} %
\usepackage{epsfig,amsmath}
\usepackage{amssymb}
\usepackage{rotate}
\usepackage{color}
\usepackage{bm}

\DeclareFontFamily{OT1}{pzc}{}
\DeclareFontShape{OT1}{pzc}{m}{it}%
            {<-> s * [1.10] pzcmi7t}{}
\DeclareMathAlphabet{\mathscr}{OT1}{pzc}%
                                {m}{it}

\newcommand{\be}{\begin{equation}}
\newcommand{\ee}{\end{equation}}
\newcommand{\bea}{\begin{eqnarray}}
\newcommand{\eea}{\end{eqnarray}}
\def\ba#1\ea{\begin{align}#1\end{align}}

\newcommand{\refeq}[1]{Eq.~(\ref{eq:#1})}          
\newcommand{\refeqs}[2]{Eqs.~(\ref{eq:#1})--(\ref{eq:#2})}          
          
\newcommand{\reffig}[1]{Fig.~\ref{fig:#1}}

\newcommand{\vs}{\nonumber\\}       
\newcommand{\refsec}[1]{Sec.~\ref{sec:#1}}          
\newcommand{\refapp}[1]{App.~\ref{app:#1}}          
 
\renewcommand{\v}[1]{\mathbf{#1}}

\newcommand{\vx}{\v{x}}

\newcommand{\vk}{\v{k}}

\newcommand{\vy}{\v{y}}

\renewcommand{\vr}{\v{r}}

\newcommand{\<}{\langle}
\renewcommand{\>}{\rangle}

\newcommand{\eps}{\varepsilon}
\renewcommand{\d}{\delta}

\newcommand{\D}{\Delta}

\newcommand{\rhob}{\overline{\rho}}

\newcommand{\Mpch}{\,h^{-1}\,{\rm Mpc}}
\newcommand{\iMpch}{\,h\,{\rm Mpc}^{-1}}

\newcommand{\Msunh}{\,h^{-1}\,M_{\odot}}

\renewcommand{\S}{\mathcal{S}}

\newcommand{\N}{\mathcal{N}}
\newcommand{\A}{\mathcal{A}}

\newcommand{\s}{\sigma}

\newcommand{\M}{\mathcal{M}}
\renewcommand{\O}{\mathcal{O}}
\newcommand{\F}{\mathcal{F}}
\newcommand{\fNL}{f_{\rm NL}}

\newcommand{\perm}{\mbox{perm.}}

\begin{document}

\title{Non-Gaussian Halo Bias Beyond the Squeezed Limit}

\author{Fabian Schmidt\footnote{Einstein fellow}}
\affiliation{Department of Astrophysical Sciences, Princeton University, Princeton, NJ 08544, USA}

\begin{abstract}
Primordial non-Gaussianity, in particular the coupling of modes with widely different
wavelengths, can have a strong impact on the large-scale clustering of tracers
through a scale-dependent bias with respect to matter.  We demonstrate that the
standard derivation of this non-Gaussian scale-dependent
bias is in general valid only in the extreme squeezed limit of the primordial bispectrum,
i.e. for clustering over very large scales.  We further show how the treatment can be generalized to describe the scale-dependent bias on smaller scales, without making any assumptions on the nature of tracers apart from a dependence on the small-scale fluctuations within a finite region.  If the leading scale-dependent bias $\D b \propto k^{\alpha}$, then the first subleading term will scale as $k^{\alpha+2}$.  This correction typically becomes relevant as one considers clustering over scales $k\gtrsim 10^{-2} \iMpch$.
\end{abstract}
\date{\today}



\maketitle

\section{Introduction}
\label{sec:intro}

Primordial non-Gaussianity is one of the most promising probes of the physics and nature of inflation in the early Universe \cite{NGReview}.  Currently the best constraints on non-Gaussianity come from the cosmic microwave background as observed by the Planck satellite \cite{PlanckNG}.  However, it has become clear recently that observations of the clustering of large-scale structure (LSS) tracers will offer competitive constraining power on non-Gaussianity \cite{MLB,coles/etal:93,luo/schramm:93,verde/heavens:01,sefusatti/komatsu:2007,slosar/etal:2008,jeong/komatsu:2009}.  

The key ingredient in describing the impact of non-Gaussianity on LSS statistics is the description of the biasing of tracers, that is, the relation to the matter density field.  \citet{dalal/etal:2008} showed that in the presence of non-Gaussianity of the local type, where the non-Gaussian potential perturbation $\phi$ is given in terms of a Gaussian field $\hat\phi$ via
\be
\phi(\vx) = \hat\phi(\vx) + \fNL\left( \hat\phi^2(\vx) - \<\hat\phi^2\> \right)\,,
\label{eq:localNG}
\ee
where $\fNL$ is a dimensionless parameter,
leads to a strongly scale-dependent bias which increases towards large scales as $k^{-2}$ in Fourier space.  That is, on large scales, tracers follow the \emph{potential}, rather than the matter as for Gaussian initial conditions.  This effect can be understood as follows.  Given a potential described by \refeq{localNG}, one can easily show \cite{dalal/etal:2008,schmidt/kamionkowski:2010} that the power spectrum of small-scale matter fluctuations in a patch of size $R_L$ around $\vx$ is rescaled as
\be
P(k_s) = [1 + 4\fNL \phi_L(\vx)] \hat P(k_s)\,,
\label{eq:uniform}
\ee
where $\phi_L(\vx)$ is the potential averaged over the patch, and $\hat P(k_s)$ is the matter power spectrum derived from the Gaussian field $\hat\phi$.  Tracers in this patch then effectively form in a Universe with higher primordial power spectrum amplitude, which will clearly change their abundance relative to other patches resulting in a modulation of the tracer density by long wavelength potential perturbations.

The rescaling \refeq{uniform} is only valid in the large-scale limit $R_L\to\infty$, that is when gradients of $\phi$ can be neglected.  As we will see, this is formally equivalent to only considering the leading contribution to the bispectrum $B_\phi$ in the squeezed limit, $\lim_{k\to 0} B_\phi(k,k_s,|\vk_s+\vk|)$.  This is clearly not a good assumption for most current large-scale structure surveys, which probe Fourier modes  $k \gtrsim 10^{-2} \iMpch$.  Several treatments have gone beyond the approximation \refeq{uniform} \cite{matarrese/verde:2008,long,scoccimarro/etal:2012}, however they all assumed specific models for the tracers (thresholding or excursion set).  It is likely that more refined models for tracers will be necessary in order to adequately describe the large samples of galaxies, quasars, and other tracers delivered by ongoing and upcoming surveys.  The goal of this paper is to go beyond the limit described by \refeq{uniform} while keeping the treatment fully independent of detailed assumptions about the tracers.

Our treatment will be based on the approach developed in \citet{PBSpaper} (see also \cite{mcdonald:2006,matsubara:2011}).  The underlying idea is that, when coarse-grained on a sufficiently large scale $R_L$, the abundance of tracers only depends locally on various coarse-grained properties of the density field.  The properties considered in \cite{PBSpaper} were the coarse-grained matter density $\rho_L = \rhob (1+\d_L)$, the curvature of the matter density $\nabla^2\rho_L$, and the amplitude of small-scale fluctuations $y_*$.  The assumption of locality is valid as long as $R_L$ is much larger than the scale of non-locality of the tracer considered.  By defining renormalized bias parameters, it is then possible to absorb the dependence on the arbitrary scale $R_L$ in the expression for tracer correlations, which then only involves large-scale matter correlators.  The renormalized bias parameters which, following conventional usage, we refer to as ``peak-background split'' (PBS) bias parameters, are given in terms of derivatives of the mean abundance of tracers with respect to the properties of the background Universe and initial conditions.  

Let us briefly recap the renormalization procedure presented in
\cite{PBSpaper} for tracer clustering in the presence of primordial non-Gaussianity.  In this paper we will assume that the leading non-Gaussian contribution is given by the bispectrum;  the generalization to higher $N$-point functions is straightforward.  If we assume that, on large scales, tracer correlations are completely described by the dependence of the tracer density on the coarse-grained fractional matter density perturbation $\d_L$, 
\be
\d_L(\vx) = \int d^3 \vy\:W_L(|\vx-\vy|) \d(\vy)\,,
\ee
where $W_L$ is an arbitrary spherically symmetric filter function on the
scale $R_L$, then the tracer correlation function $\xi_h$ is to leading order given by
\be
\xi_h(r) = b_1^2 \xi_L(r) + b_1 b_2 \< \d_L(1) \d_L^2(2)\> 
+ \cdots\,,
\label{eq:xihleading}
\ee
where $\xi_L(r) = \<\d_L(1)\d_L(2)\>$, and $b_N$ are the renormalized PBS 
bias parameters defined as derivatives of the mean tracer abundance with respect to a change in the background density $\rhob$ of the Universe.  

The leading non-Gaussian modification is the second term $\propto b_1 b_2$.  
In terms of the matter bispectrum $B_m(k_1,k_2,k_3)$, this three-point correlator
is given by
\ba
 \<\d_L(1)\d_L^2(2)\> =\:&
 \int \frac{d^3 k}{(2\pi)^3} e^{i\vk\cdot\vr} W_L(k) 
\int \frac{d^3 k_1}{(2\pi)^3} \label{eq:3ptterm}\\
 \times & W_L(k_1) W_L(|\vk+\vk_1|) 
B_m(k,k_1,|\vk+\vk_1|) \, . \nonumber
\ea
In the approach described in \cite{PBSpaper}, \refeq{xihleading} is only
a valid description of tracer correlations as long as it is independent
of the value of the coarse-graining scale $R_L$.  In order for this
to be satisfied, we need to write 
\refeq{3ptterm} in a form that is separable in $r$ and $R_L$.  For realistic
bispectra $B$, this can in general only be done approximately.  Fortunately, in the
case of primordial non-Gaussianity where the contribution in \refeq{3ptterm} typically becomes significant
for large separations $r$, we can make use of a separation of scales:  
while the Fourier integral over $k$ roughly picks out scales of $k\sim 1/r$,
the integral over $k_1$ peaks for $k_1 \sim 1/R_L$.  We thus perform
an expansion of \refeq{3ptterm} in powers of $k/k_1 \sim k R_L$.  

Let us restrict to local primordial non-Gaussianity for the time being.  As shown in \cite{PBSpaper}, to leading order in powers of $k/k_1$,
\ba
\<\d_L(1) \d_L^2(2)\> =\:& 4\fNL \s_L^2 \xi_{\phi\d}(r)\,,
\label{eq:3ptterm0}
\ea
where $\xi_{\phi\d}$ is the potential-matter cross-correlation function.  
Clearly, this contribution is strongly $R_L$-dependent through the factor $\s_L^2$.  The solution introduced in \cite{PBSpaper} is to explicitly account for the dependence of the tracer density on the local amplitude of \emph{small-scale} matter fluctuations, $y_* = (\d_s^2/\s_s^2 - 1)/2$, by generalizing the renormalized local bias parameters $b_N$ to a bivariate bias expansion $b_{NM}$ \cite{mcdonald:2008,giannantonio/porciani:2010}, where $b_{N0}$ are equal to the local biases $b_N$.  The lowest order new bias parameter $b_{01}$, which corresponds to the response of the mean tracer abundance to a change in the matter power spectrum normalization, then absorbs the term \refeq{3ptterm0}, leading to an $R_L$-independent final expression for the 2-point correlation,
\be
\xi_h(r) = b_{10}^2 \xi_L(r) + 2 b_{10} b_{01} 2 \fNL \xi_{\phi\d}(r) + \cdots\,.
\ee
The second term here corresponds to the scale-dependent bias identified in \cite{dalal/etal:2008}, where $b_{01}$ quantifies the amplitude and is given by the derivative of the tracer abundance with respect to a change in the amplitude of primordial fluctuations.

Unfortunately, \refeq{3ptterm0} is only an accurate approximation to the full expression \refeq{3ptterm} on very large scales.  Taking the expansion in $k/k_1$ to order $(k/k_1)^2$, we will show below that
\ba
\<\d_L(1) \d_L^2(2)\> =\:& 2\fNL \left[ \s_L^2 \xi_{\phi\d}(r)
+ \s_{X,L}^2 \xi_{\varphi\d}(r) \right]\,,
\label{eq:3ptterm1}
\ea
where $\varphi(\vx) = -\nabla^2 \phi(\vx)$, and the spectral moment $\s_{X,L}$ is defined below [\refeq{sfY} with \refeq{Xdef}].  The crucial point is that the second term in \refeq{3ptterm1} scales differently with $r$ and $R_L$ than the first term.  Hence, there is no hope that it will be absorbed by the single bias parameter $b_{01}$ introduced in \cite{PBSpaper}.  In fact, the physical interpretation of the two terms in \refeq{3ptterm1} is quite different: while the first term quantifies the uniform rescaling of the local small-scale fluctuations by $\phi$ [\refeq{uniform}], the second term corresponds to a scale-dependent rescaling of small-scale fluctuations by the field $\varphi(\vx)$,
\be
P(k_s) \to [1 + 4\fNL \varphi(\vx) X(k_s)] P(k_s)\,.
\ee
The first effect is naturally captured by allowing for a dependence of the tracer density on the amplitude of small-scale fluctuations.  On the other hand, to capture the effect of the second term in \refeq{3ptterm1}, we need to allow for a dependence of the tracer density on the \emph{shape} of the power spectrum of small-scale fluctuations.  As we will see, such a dependence along with the associated renormalized bias parameter is exactly what is needed to absorb the $R_L$-dependence introduced by the second term in \refeq{3ptterm1}.  

The purpose of the following sections is to make these statements rigorous.  Further, we will present all derivations for a general bispectrum of primordial non-Gaussianity, as the treatment is easily phrased to encompass the general case.

The outline of the paper is as follows. In \refsec{not}, we introduce notation and conventions used throughout the paper.  \refsec{new} describes the squeezed-limit expansion of three-point correlators for general (separable) bispectra.  The scale-dependent bias beyond the squeezed limit is derived in \refsec{bias}, while examples and numerical results are presented in \refsec{ex}.  We make the connection to previous results in \refsec{thr}, and discuss other sources that become relevant in this regime in \refsec{other}.  We conclude in \refsec{concl}.

\section{Notation}
\label{sec:not}

We assume that the primordial $N$-point functions are given in terms of the Bardeen potential during matter domination, $\phi$.  Throughout, we will only deal with the statistics of $\phi$ and the initial (linear) density field, scaled to some redshift $z$.  Note in particular that the density bias parameters correspondingly denote \emph{Lagrangian} biases throughout.  The relation between $\phi$ and the linear density field at redshift $z$ is written in Fourier space as
\ba
\delta(\vk,z) =\:& \M(k,z)\phi(\vk) \vs
\M(k,z) =\:& \frac{2}{3} \frac{k^2 T(k) g(z)}{\Omega_{m} H_0^2 (1+z)}\,,
\label{eq:M}
\ea
where $T(k)$ is the matter transfer function normalized to unity as $k\to 0$, 
and $g(z)$ is the linear growth rate of the gravitational potential normalized 
to unity during the matter dominated epoch.  In the following, we will drop the argument $z$ in $\d$ and $\M$ since it is not of relevance in the derivation.  
Further, we define $\M_Y(k) = \M(k) \tilde W_Y(k)$ where $Y$ stands for different filters such as $L, s, *$ which we will encounter below.  We let $P_\phi(k)$ denote the power spectrum of $\phi$, and $P_m(k) = \M^2(k) P_\phi(k)$ the matter power spectrum (again dropping the argument $z$).  

We will need various spectral moments of the density field.  We define
\be
\s^2_Y = \int \frac{d^3 k}{(2\pi)^3} P_m(k) \tilde W_Y^2(k)
\label{eq:sY}
\ee
for any filter $W_Y$.  Further, given a weighting function $f(k)$, we define
\be
\s^2_{f,Y} = \int \frac{d^3 k}{(2\pi)^3} f(k) P_m(k) \tilde W_Y^2(k)\,.
\label{eq:sfY}
\ee
It will also be useful to define non-local transformations of the density field.  Again given a function $f(k)$, we let
\be
\d_{f,Y}(\vx) = \int \frac{d^3k}{(2\pi)^3} e^{i \vk\vx} f(k) \tilde W_Y(k) \d(\vk)\,.
\label{eq:dfY}
\ee
In particular, this yields
\be
\< \d_{f, Y} \d_Y \> = \s^2_{f,Y}\,.
\ee

\section{Beyond the squeezed limit}
\label{sec:new}

We begin with deriving the correlator \refeq{3ptterm} and expanding it in the squeezed limit.  More details are provided in \refapp{corrderivN}.  Throughout, we will work to
leading order in the dimensionless amplitude of non-Gaussianity $\fNL$.  
At this order and for the models we consider, the only relevant $N$-point 
functions are the power spectrum and bispectrum.  

Expressed in terms of the bispectrum of primordial perturbations $B_\phi$, defined through
\be
\< \phi_{\vk}\phi_{\vk_1}\phi_{\vk_2}\>
= (2\pi)^3 \d_D(\vk+\vk_1+\vk_2) B_\phi(\vk,\vk_1,\vk_2)\,,
\ee
the term we are interested in becomes
\ba
 \<\d_L(1)\d_L^2(2)\> =\:&
 \int \frac{d^3 k}{(2\pi)^3} e^{i\vk\cdot\vr} \M_L(k) 
\int \frac{d^3 k_1}{(2\pi)^3} 
\M_L(k_1) \vs
& \times \M_L(|\vk+\vk_1|) 
B_\phi(k,k_1,|\vk+\vk_1|) \, . 
\label{eq:diffterms}
\ea
Our goal is to obtain an expression that is separable in $r$ and $R_L$.  
Neglecting the $R_L$-dependence introduced through $\M_L(k)$, which is irrelevant if $r \gg R_L$ (see \refsec{other}), separability in $r$ and $R_L$ is equivalent to having an integrand separable in $\vk$ and $\vk_1$.  As a necessary prerequisite, we assume that the bispectrum $B_\phi$ is given in separable form,
\ba
B_\phi(k_1, k_2, k_3) = \sum_\alpha A_\alpha \Big[& F^{(1)}_\alpha(k_1) F^{(2)}_\alpha(k_2) F^{(3)}_\alpha(k_3) \vs
& + 5\:{\rm perm.} \Big]\,,
\label{eq:Bphigen}
\ea
where $A_\alpha$ are constants and the 6 permutations guarantee the symmetry of $B_\phi$ in its arguments.  The sum $\alpha$ runs over however many terms are necessary to adequately approximate the bispectrum in separable form.  We further define 
\be
\tilde F^{(i)}_\alpha(k) = \M_L(k) F^{(i)}_\alpha(k)\,.
\ee
\refeq{diffterms} then becomes
\ba
&  \<\d_L(1)\d_L^2(2)\> = \sum_\alpha A_\alpha\bigg\{\int \frac{d^3 k}{(2\pi)^3} \tilde F^{(1)}(k) e^{i\vk\cdot\vr} \label{eq:diffterms2}\\
& \quad\times\int \frac{d^3 k_1}{(2\pi)^3} \tilde F^{(2)}_\alpha(k_1) \tilde F^{(3)}_\alpha(|\vk_1 + \vk|) 
+ 5\:{\rm perm.} \bigg\}\,.
\nonumber
\ea
We now expand the $k_1$ integrand in powers of $k/k_1$ up to second order.  As shown in \refapp{corrderivN},
\begin{widetext}
\ba
 \<\d_L(1)\d_L^2(2)\> = \sum_\alpha A_\alpha \bigg\{& \int \frac{d^3 k}{(2\pi)^3} \tilde F^{(1)}_\alpha(k) e^{i\vk\cdot\vr} \vs
\times& \int \frac{d^3 k_1}{(2\pi)^3} 
\bigg[2 \tilde F_\alpha^{(2)}(k_1)\tilde F_\alpha^{(3)}(k_1) 
+ \frac12  \frac{k^2}{k_1^2} 
\tilde F_\alpha^{(2)}(k_1)\left[(1-\mu^2) k_1 \tilde F'^{(3)}_\alpha(k_1) + \mu^2 k_1^2 \tilde F''^{(3)}_\alpha(k_1)\right]
\vs
 +& \frac12  \frac{k^2}{k_1^2} 
\tilde F_\alpha^{(3)}(k_1)\left[(1-\mu^2) k_1 \tilde F'^{(2)}_\alpha(k_1) + \mu^2 k_1^2 \tilde F''^{(2)}_\alpha(k_1)\right]
\bigg]  + \{ (123) \to (231) \} + \{ (123) \to (312) \}
\bigg\} 
\,,\nonumber
\ea
\end{widetext}
where primes denote derivatives with respect to $k_1$.  This expression is now in the desired separable form.  It is valid up to terms of order $k^4/k_1^4$ as the cubic terms drop out just like the linear terms did.   We can make this result more obvious and compact by introducing the notation
\ba
\xi^{(i)}_{\alpha L}(r) \equiv\:& \int \frac{d^3 k}{(2\pi)^3} \M_L(k) F^{(i)}_\alpha(k) e^{i\vk\cdot\vr} 
\label{eq:xialpha}\\
\xi^{(i)}_{\nabla^2\alpha L}(r) \equiv\:& \int \frac{d^3 k}{(2\pi)^3} \M_L(k) k^2\,F^{(i)}_\alpha(k) e^{i\vk\cdot\vr}
\label{eq:xinalpha}\\
\s^{2(ij)}_{\alpha L} \equiv\:& \frac1{2\pi^2} \int_0^\infty dk_1\,k_1^2 \M_L^2(k_1)
F_\alpha^{(i)}(k_1) F_\alpha^{(j)}(k_1)
\label{eq:sigalpha}\\
\s^{2(ij)}_{X\alpha L} \equiv\:& \frac1{12\pi^2} \int_0^\infty dk_1 
\Big\{\tilde F_\alpha^{(i)}(k_1) \vs
& \hspace*{2.1cm}\times \left[2 k_1 \tilde F'^{(j)}_\alpha(k_1) + k_1^2 \tilde F''^{(j)}_\alpha(k_1)\right] \vs
& \hspace*{2cm} + (i) \leftrightarrow (j) \Big\}\,.
\label{eq:sigXalpha}
\ea
Note that for all spectral moments, $\s^{2(ij)} = \s^{2(ji)}$, and that the derivatives in \refeq{sigXalpha} act on both $F^{(i)}_\alpha(k_1)$ and $\M_L(k_1)$.  With this, we obtain
\ba
 \<\d_L(1)\d_L^2(2)\>=&\sum_\alpha A_\alpha\bigg\{ 2\xi^{(1)}_{\alpha L}(r) \s^{2(23)}_{\alpha L}
+ \xi^{(1)}_{\nabla^2\alpha L}(r) \s^{2(23)}_{X \alpha L} \vs
& + \{ (123) \to (231) \} + \{ (123) \to (312) \}
\bigg\}\,,
\label{eq:diffterms_gen}
\ea
where the second line denotes the two remaining cyclic permutations.  Apart
from the residual $R_L$-dependence in $\xi_{\alpha L},\:\xi_{\nabla^2\alpha L}$, this
expression is fully separable in $r$ and $R_L$ as desired.  The expansion in $k/k_1$ can of course also be taken to higher order if necessary.  We will discuss the necessity of this below.  

Given the appearance of $\s_{\alpha L}^2$ and $\s_{X \alpha L}^2$ in \refeq{diffterms_loc}, 
the non-Gaussian correction to $\xi_h$ is strongly $R_L$-dependent.  In the next section, we will address this issue through renormalized bias parameters.  

\subsection{Primordial non-Gaussianity of the local type}
\label{sec:localNG}

\refeq{diffterms_gen} is general, but somewhat abstract.  In order to
clarify its physical significance, we now specialize to the case of local 
primordial non-Gaussianity, for which the bispectrum is given by
\ba
B_\phi(\vk_1,\vk_2,\vk_3) =\:& 2\fNL [P_\phi(k_1)P_\phi(k_2) + 2\:\perm]\,.
\label{eq:Bphiloc}
\ea
Since $B_\phi$ is already separable, we only have one term in the sum \refeq{Bphigen}, with $A_\alpha = A = \fNL$, and 
\be
F^{(1)}(k) = F^{(2)}(k) = P_\phi(k);\quad F^{(3)}(k) = 1\,.
\ee 
The permutation $\xi_{\alpha L}^{(3)}(r) \s^{2(12)}_{\alpha L}$ in \refeq{diffterms_gen} is suppressed (for scale-invariant $P_\phi$) by $(k/k_1)^3$ relative to the other two identical permutations.  In keeping with our treatment up to $(k/k_1)^4$, we will retain the leading contribution from this term.  We then obtain
\ba
 \<\d_L(1)\d_L^2(2)\> =\:& 
4\fNL \s_L^2 \xi_{\phi\d_L}(r) + 2\fNL \s_{X,L}^2 \xi_{\varphi\d_L}(r) \vs
& + 2 \fNL \s_{\alpha}^{2(12)} \xi_{\alpha L}^{(3)}(r) \,,
\label{eq:diffterms_loc}
\ea
where we have defined
\ba
\xi_{\phi\d_L}(r) =\:& \frac{d^3 k}{(2\pi)^3} \M_L(k) P_\phi(k) e^{i\vk\cdot\vr} = \<\phi(1) \d_L(2)\>
\label{eq:locdefs}\\
\xi_{\varphi\d_L}(r) =\:& \frac{d^3 k}{(2\pi)^3} \M_L(k) k^2 P_\phi(k) e^{i\vk\cdot\vr} = \<\varphi(1) \d_L(2)\>
\vs
X(k) =\:& \frac1{6k^2} \Big \{\M_L^{-1}(k) \left[2 k \M'_L(k) + k^2 \M''_L(k)\right] \label{eq:Xdef}\\
&+ \left( P_\phi \M_L\right)^{-1}_k \left[2 k (P_\phi \M_L)'_{k} + k^2 (P_\phi\M_L)''_{k}\right] 
\Big\}\,, \nonumber
\ea
and $\s_L^2,\,\s_{X,L}^2$ are defined through \refeqs{sY}{sfY}.  Here we have introduced $\varphi(\vx) \equiv -\nabla^2 \phi(\vx)$.  The contribution from the third permutation is given by 
\ba
\xi^{(3)}_{\alpha L}(r) \equiv\:& \int \frac{d^3 k}{(2\pi)^3} \M_L(k) e^{i\vk\cdot\vr} \vs
\s^{2(12)}_{\alpha L} \equiv\:& \frac1{2\pi^2} \int_0^\infty dk_1\,k_1^2 P_\phi(k_1) P_m(k_1) \tilde W_L^2(k_1)\,.\label{eq:thirdloc}
\ea

The first (leading) term in \refeq{diffterms_loc} is well known and agrees with that derived in \cite{PBSpaper}.  This term can effectively be described as a rescaling of the local density field,
\be
\d(\vx) \to \left[1 + 2\fNL \phi(\vx)\right] \d(\vx)\,,
\ee
which leads to \refeq{uniform}.  On the other hand, the second (subleading) term in \refeq{diffterms_loc} can be seen as coupling the Laplacian of $\phi$ to the density field.  However, this coupling is scale-dependent, i.e. when performing a Fourier transform in a local patch where $\varphi$ can be considered constant, we have
\ba
\d(\vk) \to\:& \left[1 + \fNL X(k) \varphi(\vx)\right] \d(\vk) \,.
\ea
This effect is of course suppressed with respect to $\phi$ in the large-scale limit.  In Fourier space on large scales
$\M(k) P_\phi(k) \propto k^{-2} P_m(k)$, so that $\xi_{\phi\d}(r)$ grows with
respect to $\xi_L(r)$ on large scales, while $\xi_{\varphi\d,L}(r) \sim \xi_L(r)$ 
(the nontrivial transfer function leads to departures on scales smaller than
about $100\Mpch$).  

\section{Scale-dependent Bias}
\label{sec:bias}

As shown in \cite{PBSpaper}, one can absorb the leading term $\propto \s_L^2$ 
in \refeq{diffterms_loc} by introducing
a dependence of the tracer density on the local variance of the small-scale
density field.  We define the small-scale density field as the local 
fluctuations around the coarse-grained field $\d_L$:
\ba
\d_s(\vx) \equiv\:& \d_*(\vx) - \d_L(\vx) \label{eq:dsdef}\\
=\:& \int d^3\vy [W_*(\vx-\vy) - W_L(\vx-\vy)] \d(\vy) \vs
=\:& \int \frac{d^3\vk}{(2\pi)^3} \tilde W_s(k) \d(\vk)
e^{i\vk \cdot \vx}, \vs
\tilde W_s(k) =\:& \tilde W_*(k) - \tilde W_L(k)\,,
\label{eq:Ws}
\ea
where we have introduced a fixed small smoothing scale $R_*$.  Thus, we write
\ba
n_h[\d_L(\vx)] \to\:& n_h[\d_L(\vx), y_*(\vx)] \label{eq:nh}\\
y_*(\vx) \equiv\:& \frac12 \left(\frac{\d_s^2(\vx)}{\s_s^2} - 1\right) \, ,
\label{eq:ydef}
\ea
where the subscript $*$ refers to the smoothing scale $R_*$, $\<y_*\> =0$, 
and the factor of $1/2$ is included to obtain expressions which conform to 
standard convention.  A very similar
derivation to that of \refeq{diffterms_gen} yields (\refapp{corrderivN})
\ba
 \<\d_L(1) y_*(2)\> =\sum_\alpha A_\alpha\bigg\{& \xi^{(1)}_{\alpha L}(r) \frac{\s^{2(23)}_{\alpha s}}{\s_s^2} 
+ \frac12 \xi^{(1)}_{\nabla^2\alpha L}(r) \frac{\s^{2(23)}_{X \alpha s}}{\s^2_{s}} \vs
& + 2\:{\rm perm.}
\bigg\}\,,
\label{eq:dy_gen}
\ea
where a subscript $s$ indicates that the filter function $\tilde W_L$ should be replaced with $\tilde W_s$ in \refeqs{sigalpha}{sigXalpha}.  
In the case of local primordial non-Gaussianity, this again reduces to
\ba
 \<\d_L(1) y_*(2)\> = 2\fNL \bigg\{& \xi_{\phi \d_L}(r)  
+ \frac12 \xi_{\varphi \d_L}(r) \frac{\s^2_{X s}}{\s^2_{s}} \vs
& + \frac12 \xi^{(3)}_{\alpha L} \frac{\s^{2(12)}_{\alpha s}}{\s^2_s}
\bigg\}\,.
\label{eq:dy_loc}
\ea
While the renormalization of the $R_L$-dependence in \refeq{diffterms_gen} works for the leading term as will see now, we already notice two issues with \refeq{dy_loc} in comparison with \refeq{diffterms_loc}.  First, the second and third term in \refeq{dy_loc} depend on $R_L$, since $\tilde W_s$ is defined with respect to $\tilde W_L$ [\refeq{Ws}].  Thus, the final expression cannot simply involve $\< \d_L(1) y_*(2)\>$.  Second, there is no reason why the relative magnitude of the three terms in \refeq{dy_loc} should be the same as for the three terms in \refeq{diffterms_loc}, so that we cannot expect all terms in \refeq{diffterms_loc} to be absorbed by $\<\d_L(1) y_*(2)\>$.  These issues will be resolved below.  

\subsection{Bivariate bias expansion}
\label{sec:bi}

We begin by re-examining the bivariate bias expansion of \cite{PBSpaper}
(which in the local case at lowest order is equivalent to what was presented in previous papers \cite{mcdonald:2008,giannantonio/porciani:2010}).  Throughout this section and the next, we will restrict to a single separable contribution $\alpha$ and a single permutation $(123)$ of \refeq{diffterms_gen} [note that this corresponds to two permutations in \refeq{Bphigen}].  The complete prescription for tracer clustering, presented in \refsec{sum} below, will then involve a sum over the contributions from different $\alpha$ and permutations.  

In terms of the ``bare'' bias parameters $c_{nm}$, that is, the coefficients of the Taylor series of $n_h$ in $\d_L,\,y_*$, the tree-level expression for the tracer correlation function in the bivariate PBS expansion is [see Eq.~(109) in \cite{PBSpaper}]
\ba
& \xi_h^{\rm bare}(r) = \frac1{\N^2}\bigg\{ c_{10}^2 \xi_L(r) + c_{10} c_{20} \<\d_L(1) \d_L^2(2)\> \vs
& \hspace*{3cm} + 2 c_{10} c_{01} \<\d_L(1) y_*(2)\> \bigg\} \vs
& = \frac1{\N^2}\bigg\{  c_{10}^2 \xi_L(r) \label{eq:xihbare}\\
& \hspace*{1cm} + c_{10} c_{20} A_\alpha\bigg[  2\xi^{(1)}_{\alpha L}(r) \s^{2(23)}_{\alpha L} 
+ \xi^{(1)}_{\nabla^2\alpha L}(r) \s^{2(23)}_{X \alpha L} \bigg]
\vs
& \hspace*{1cm} + 2 c_{10} c_{01} A_\alpha\bigg[ \xi^{(1)}_{\alpha L}(r) \frac{\s^{2(23)}_{\alpha s}}{\s_s^2} 
+ \frac12 \xi^{(1)}_{\nabla^2\alpha L}(r) \frac{\s^{2(23)}_{X \alpha s}}{\s^2_{s}}\bigg]
\bigg\} 
\nonumber
\ea
where
\be
\N \equiv \sum_{n,m=0}^\infty \frac{c_{nm}}{n!m!} \<\d_L^n y_*^m\>\,.
\label{eq:NNG}
\ee
Let us focus on the leading terms in the squeezed limit first.  We 
need to generalize the definition of the renormalized bias parameters $b_{NM}$ to the case of the general separable bispectrum \refeq{Bphigen}.  

Consider a rescaling of the Fourier-space density field given by
\ba
\d(\vk_1) \to\:& \d(\vk_1)\left[1 + \eps F(k_1) \right] \vs
F(k_1) =\:& \frac{F^{(2)}_\alpha(k_1) F^{(3)}_\alpha(k_1)}{P_\phi(k_1)}\,.
\label{eq:epstrans} 
\ea
Further, we include a shift in the density perturbation $\d \to \d+D$, corresponding to adding a uniform matter density of $D\rhob$.  
We then define the bivariate bias parameters $b_{NM}$ as the response of the mean tracer density to this uniform matter density and the rescaling of the initial density field under \refeq{epstrans} (both to be evaluated at fixed proper time):
\be
b_{NM} \equiv \frac{1}{\<n_h\>_{D=0,\eps=0}} \frac{\partial^{N+M} \<n_h\>_{D,\eps}}{\partial D^N \partial \eps^M}\bigg|_{D=0,\eps=0}.
\label{eq:bNM}
\ee
Note that $b_{NM}$ should really be denoted $b_{NM}^{\alpha (23)}$ here, since the definition refers to the specific rescaling in \refeq{epstrans}.  However, since we are only dealing with a single $\alpha$ and permutation here and in \refsec{tri}, we will drop this designation for notational clarity.  Note further that the dimension of $b_{01}$ is such that the contribution to the correlation function is dimensionless.  In particular,
\ba
{\rm dim}(b_{01}) =\:& {\rm dim}\left[ F(k_1) \right] 
= \frac{{\rm Mpc}^3}{{\rm dim}(F_\alpha^{(1)})}\,,
\label{eq:dimb01}
\ea
where we have assumed that the $A_\alpha$ are dimensionless so that ${\rm dim}(F^{(1)}_\alpha F^{(2)}_\alpha F^{(3)}_\alpha) = {\rm Mpc}^6$ [\refeq{Bphigen}].  Under the rescaling \refeq{epstrans}, we have [using the definition \refeq{dfY}]
\ba
\d_L(\vx) \to\:& \d_L(\vx) + \eps\: \d_{F,L}(\vx) \vs
y_*(\vx) \to\:& y_*(\vx) + \frac{\eps}{\s_s^2} \d_s(\vx) \d_{F,s}(\vx)
+ \frac{\eps^2}{2\s_s^2} \d_{F,s}^2(\vx)
\; .
\ea
Note that
\be
\< \d_Y(\vx) \d_{F,Y}(\vx)\> = \s_{F,Y}^2 = \s^{2 (23)}_{\alpha Y}
\label{eq:dFy}
\ee
for $Y = L,\,s,\,\dots\,$.  Using \refeq{nh}, we obtain
\ba
\<n_h\>(D,\eps) =\:& \<n_h(0)\> \sum_{n,m} \frac{c_{nm}}{n!m!} \vs
& \times \Bigg\< \Big[(\d_L + \eps\, \d_{F,L})\d_L+D\Big]^n \vs
& \hspace*{0.7cm} \times
\left [y_* + \frac{\eps}{\s_s^2} \d_s \d_{F,s} + \frac{\eps^2}{\s_s^2} \d_{F,s}^2 \right]^m \Bigg\>.\nonumber
\ea
To the order we are interested in, we only need $b_{01}$, which is given by
\ba
b_{01} =\:& \frac1{\N} \sum_{n,m} \frac{c_{nm}}{n!m!} \vs
&\quad\times \left(n \< \d_{F,L} \d_L^{n-1} y_*^m\> + \frac{m}{\s_s^2} 
\< \d_L^n  \d_s \d_{F,s} y_*^{m-1}\Big\>\right)
\vs
=\:& \frac1{\N} \left( c_{01} \frac{\s^{2(23)}_{\alpha s}}{\s_s^2}
+ c_{20} \sigma^{2(23)}_{\alpha L} + \O(\d^3, \fNL^2) \right).
\label{eq:b01}
\ea
Following the same reasoning as in \cite{PBSpaper}, our guess for the tree-level tracer
correlation function written in terms of renormalized bias parameters is
\ba
\xi_h^{\rm renorm}(r) =\:& b_{10}^2 \xi_L(r) + 2 b_{10} b_{01} \frac{\s_s^2}{\s^{2(23)}_{\alpha s}} \< \d_L(1) y_*(2) \> 
\vs
=\:& b_{10}^2 \xi_L(r) \label{eq:xihrenorm1}\\
&+ 2 b_{10} b_{01} A_\alpha \bigg\{
\xi^{(1)}_{\alpha L}(r) 
+ \frac12 \xi^{(1)}_{\nabla^2\alpha L}(r) \frac{\s^{2(23)}_{X \alpha s}}{\s^{2 (23)}_{\alpha s}} \bigg\}
\,, \nonumber
\ea
where the prefactor in front of $\<\d_L(1) y_*(2)\>$ takes into account that
the tree-level relation between $b_{01}$ and $c_{01}$ is $b_{01} = c_{01} \s^{2(23)}_{\alpha s} / \s_s^2$.  In the second line we have used \refeq{dy_gen}.  The fact that renormalization is not successful at subleading order can already be seen from the last term in this equation.  This depends on $R_L$ through the spectral moments, since the kernel $\tilde W_s$ depends on $R_L$.  Thus, we assume that a proper renormalization will absorb this term as well, and write our updated guess as
\ba
\xi_h^{\rm renorm}(r) =\:& b_{10}^2 \xi_L(r) 
+ 2 b_{10} b_{01} A_\alpha \xi^{(1)}_{\alpha L}(r) 
\,. \label{eq:xihrenorm2}
\ea
Inserting the expression for $b_{01}$ [\refeq{b01}] yields
\ba
\xi_h^{\rm renorm}(r) = \frac1{\N^2} \Bigg\{& c_{10}^2 \xi_L(r) 
+ 2 c_{10} c_{01} \frac{\s^{2(23)}_{\alpha s}}{\s_s^2} A_\alpha 
\xi^{(1)}_{\alpha L}(r) \vs
& + 2 c_{10} c_{20} \s^{2(23)}_{\alpha L} A_\alpha 
\xi^{(1)}_{\alpha L}(r) \Bigg\} \,.
\label{eq:xihrenorm3}
\ea
If the renormalized bias $b_{01}$ defined with respect to the transformation \refeq{epstrans} properly removes all $R_L$-dependence of the tracer 2-point function, \refeq{xihrenorm2} should match the bare bias expansion \refeq{xihbare}.  Let us thus take the difference:
\ba
\xi_h^{\rm bare}(r) -&\: \xi_h^{\rm renorm}(r) 
\label{eq:remainders}\\
& = A_\alpha \xi^{(1)}_{\nabla^2\alpha L}(r)  c_{10} \Bigg[
 c_{20} \s^{2(23)}_{X \alpha L} 
+  c_{01} \frac{\s^{2(23)}_{X \alpha s}}{\s_s^2} \Bigg]\,.
\nonumber
\ea
All terms here are sub-leading in the squeezed limit, that is, at leading order in the squeezed limit the renormalization works in the same way as shown in \cite{PBSpaper}.  However, when going beyond the squeezed limit for general primordial non-Gaussianity, the description of the tracer density as a bivariate function $n_h[\d_L, y_*]$ is not sufficient.

\subsection{Trivariate bias expansion}
\label{sec:tri}

We are seeking a third local parameter (besides $\d_L$ and $y_*$) which
the tracer density in general depends on, and which will allow us to absorb the $R_L$-dependent terms in \refeq{remainders}.  In the local model, the residual $R_L$-dependence that we encounter when going beyond the squeezed limit is induced by the fact that in this case, the effect of long-wavelength modes on small-scale modes is not simply to rescale them uniformly, but to also change the shape of the small-scale power spectrum [\refeq{Xdef}].  More generally, small-scale modes are rescaled differently as function of the correlation scale $r$, as the relative importance of the terms given by $\xi^{(1)}_{\alpha L}(r)$ and $\xi^{(1)}_{\nabla^2\alpha L}(r)$ changes.  Thus, we have to allow for a dependence of the tracer density not only on the local variance of small-scale fluctuations, but also on their shape.  As a proxy for the power spectrum shape, we will use
\ba
\mu_*(\vx) \equiv\:& \frac{d}{d\ln R_*} y_*(\vx)\,.
\label{eq:mudef}
\ea
This choice is useful because transformations of the density field of the form
\be
\d_L \to \d_L + D; \quad \d(\vx) \to \left[1 + \iota\right] \d(\vx)
\ee
leave $\mu_*(\vx)$ invariant.  Using \refeq{mudef} and \refeq{dy_gen} we obtain, again for one of the three cyclic permutation of the contribution $\alpha$,
\ba
\< \d_L(1) \mu_*(2)\> =\:& A_\alpha\bigg\{ \xi^{(1)}_{\alpha L}(r) 
\frac{d}{d\ln R_*} \left(\frac{\s^{2(23)}_{\alpha s}}{\s_s^2} \right) \vs
& + \frac12 \xi^{(1)}_{\nabla^2\alpha L}(r) \frac{d}{d\ln R_*} \left(\frac{\s^{2(23)}_{X \alpha s} }{\s_s^2} \right)
\bigg\}\,.
\label{eq:mucorr}
\ea
For local non-Gaussianity, $\s^2_{\alpha s} = \s^2_s$, so that the first term drops out, and we obtain (restoring the number of permutations)
\ba
\< \d_L(1) \mu_*(2)\> \stackrel{\rm local}{=}\:& 4\fNL \frac12 \xi^{(1)}_{\varphi L}(r) \frac{d}{d\ln R_*} \left(\frac{\s^{2}_{X s} }{\s_s^2} \right) \vs
=\:& 2\fNL \xi^{(1)}_{\varphi L}(r)  \frac{\s^{2}_{X s}}{\s_s^2} 
\left(\frac{d\ln \s^2_{X s}}{d\ln R_*} - \frac{d \ln \s^2_{s}}{d\ln R_*} \right) \,.
\nonumber
\ea
Let us now adopt $\mu_*(\vx)$ as third local parameter:
\be
n_h\left[\d_L(\vx), y_*(\vx)\right] \to n_h\left[\d_L(\vx), y_*(\vx), \mu_*(\vx)\right]\,.
\ee
We again consider the scale-dependent transformation \refeq{epstrans}, and in addition a further, independent scale-dependent transformation of the density field
\be
\d(\vk) \to \left[1 + \iota f(k) \right] \d(\vk)\,.
\label{eq:iotatrans}
\ee
Here we leave the function $f(k)$ free for the moment.  This leads to
\ba
\d_L(\vx) \to\:& \d_L(\vx) + \iota \d_{f,L}(\vx) \vs
y_*(\vx) \to\:& y_*(\vx) + \frac{\iota}{\s_s^2} \d_s(\vx) \d_{f,s}(\vx)
+ \frac{\iota^2}{\s_s^2} \d_{f,s}^2(\vx) 
\label{eq:mutrans} \\
\mu_*(\vx) \to\:& \mu_*(\vx) + \frac{d}{d\ln R_*} \left[
\frac{\iota}{\s_s^2} \d_s(\vx) \d_{f,s}(\vx)
+ \frac{\iota^2}{\s_s^2} \d_{f,s}^2(\vx) \right]\,.
\nonumber
\ea
In keeping with the leading order treatment in $\fNL$, we will only consider
the first derivative of the mean tracer density with respect to $\iota$.  Thus,
we can drop the $\iota^2$ terms in \refeq{mutrans}.  We define
\be
b_{NML} = \frac1{\< n_h(0) \>} \frac{\partial^{N+M+L} \< n_h(D,\eps,\iota)\>}{\partial D^N \eps^M \iota^L}\Big|_0\,,
\ee
where the mean tracer number density is given by
\begin{widetext}
\ba
\< n_h \> (D,\eps, \iota) = \<n_h(0)\> \sum_{nml} \frac{c_{nml}}{n!m!l!}
\Bigg\< & \left[\left(1 + \eps F(k) + \iota f(k)\right) \d_L + D \right]^n
\left[y_* + \eps \frac{\d_s \d_{F,s}}{\s_{s}^2} + \iota \frac{\d_s \d_{f,s}}{\s_s^2} \right]^m \vs
& \times
\left[\mu_* + \eps \frac{d}{d\ln R_*}\left(\frac{\d_s \d_{F,s}}{\s_s^2}
\right) + \iota \frac{d}{d\ln R_*}\left(\frac{\d_s \d_{f,s}}{\s_s^2} \right)
\right]^l \Bigg\>\,.
\ea
\end{widetext}
We thus obtain 
\ba
b_{N00} =\:& b_{N0} \vs
b_{010} =\:& \frac1{\<n_h\>(0)} \frac{\partial \<n_h\>(D,\iota,\eps)}{\partial\eps} \Big|_0 \label{eq:b010}\\
%
%
%
%
=\:& \frac1{\N} \left[ c_{200} \s_{F,L}^2
+ c_{010} \frac{\s^2_{F,s}}{\s_{s}^2} 
+ c_{001} \frac{d}{d\ln R_*}\left(\frac{\s^2_{F,s}}{\s_s^2} \right)
\right]
\,,
\nonumber
\ea
and
\ba
b_{001} =\:& \frac1{\<n_h\>(0)} \frac{\partial \<n_h\>(D,\iota,\eps)}{\partial\iota} \Big|_0 
\label{eq:b001} \\
=\:& \frac1{\N} \left[c_{200} \s^2_{f,L} + c_{010} \frac{\s_{f,s}^2}{\s_s^2}
+ c_{001} \frac{d}{d\ln R_*}\left(\frac{\s^2_{f,s}}{\s_s^2} \right)
\right]\,,
\nonumber
\ea
where we have expanded to order $\d_L^2$.  Note the non-trivial coefficient multiplying $c_{001}$ in \refeq{b001}.  We thus have to 
divide $b_{001}$ by this coefficient when writing down the contributions to 
$\xi_h$ that correlate $\mu_*$ with other perturbations.  

We now update our ansatz for the renormalized tracer correlation [\refeq{xihrenorm2}] to the trivariate case.  Since we found in the previous section that the bivariate case (neglecting the dependence of $n_h$ on $\mu_*$) successfully produced an $R_L$-independent expression for the leading squeezed-limit contribution, we expect that only the sub-leading contribution of $\<\d_L(1)\mu_*(2)\>$ [\refeq{mucorr}] will contribute.  That is, we expect the leading contribution in \refeq{mucorr} to be absorbed by $b_{010}$.  Thus, our expectation is
\ba
\xi_h^{\rm tri}(r) =\:& b_{100}^2 \xi_L(r) 
+ 2 b_{100} b_{010} A_\alpha \xi^{(1)}_{\alpha L}(r) 
\label{eq:xihtri}\\
+ & 2 b_{100} b_{001} A_\alpha  \frac12 \xi^{(1)}_{\nabla^2\alpha L}(r)
\frac{d\left(\s^{2(23)}_{X \alpha s}/\s_s^2 \right)/ d\ln R_*}
{d\left(\s^2_{f,s}/\s^s_s\right)/d\ln R_*}\,.
\nonumber
\ea
The last term is in general only guaranteed to be $R_L$-independent if 
$\s^{2(23)}_{X \alpha s} = \s^2_{f,s}$, that is, if
\ba
k_1^2 P_m(k_1) f(k_1) = \:&
\frac16 \tilde F_\alpha^{(2)}(k_1) \left[2 k_1 \tilde F'^{(3)}_\alpha(k_1) + k_1^2 \tilde F''^{(3)}_\alpha(k_1)\right]\vs
&  + (2) \leftrightarrow (3) \,.
\label{eq:fdef}
\ea
Thus, we will fix this as our choice of $f(k)$, and in the following derivation set $\s^{2(23)}_{X \alpha Y} = \s^2_{f,Y}$, where $Y= s, L$.  Note that the dimension of $b_{001}$ is equal to the dimension of $f$ and is given by
\ba
{\rm dim}(b_{001}) =\:& {\rm dim}\left(\frac{F^{(2)}_\alpha(k_1) F^{(3)}_\alpha(k_1)}{ k_1^2 P_m(k_1)}\right) 
%
= \frac{{\rm Mpc}^5}{{\rm dim}(F_\alpha^{(1)})}\,,
\label{eq:dimb001}
\ea
that is, the dimension of $b_{01}$ times Mpc$^2$.  The bare expansion then becomes [using $\s^{2(23)}_{\alpha Y} = \s^2_{F,Y}$ from \refeq{dFy}]
\begin{widetext}
\ba
 \xi_h^{\rm bare}(r) = \frac1{\N^2}\bigg\{& c_{100}^2 \xi_L(r) + c_{100} c_{200} \<\d_L(1) \d_L^2(2)\> + 2 c_{100} c_{010} \<\d_L(1) y_*(2)\> 
+ 2 c_{100} c_{001} \<\d_L(1) \mu_*(2)\> \bigg\} \vs
= \frac1{\N^2}\Bigg\{&  c_{100}^2 \xi_L(r) + c_{100} c_{200} A_\alpha\bigg[  2\xi^{(1)}_{\alpha L}(r) \s^{2}_{F, L} 
+ \xi^{(1)}_{\nabla^2\alpha L}(r) \s^{2}_{f, L} \bigg]
\vs
& + 2 c_{100} c_{010} A_\alpha\bigg[ \xi^{(1)}_{\alpha L}(r) \frac{\s^{2}_{F, s}}{\s_s^2} 
+ \frac12 \xi^{(1)}_{\nabla^2\alpha L}(r) \frac{\s^{2}_{f, s}}{\s^2_{s}}\bigg] \vs
& + 2 c_{100} c_{001} A_\alpha\bigg[ \xi^{(1)}_{\alpha L}(r) 
\frac{d}{d\ln R_*} \left(\frac{\s^{2}_{F, s}}{\s_s^2} \right)
+ \frac12 \xi^{(1)}_{\nabla^2\alpha L}(r) \frac{d}{d\ln R_*} \left(\frac{\s^{2}_{f, s} }{\s_s^2} \right)
\bigg]\Bigg\} \vs
= \frac1{\N^2}\Bigg\{&  c_{100}^2 \xi_L(r) +
2 A_\alpha \xi^{(1)}_{\alpha L}(r) c_{100} \bigg[
 c_{200} \s^{2}_{F, L} + c_{010} \frac{\s^{2}_{F, s}}{\s_s^2} 
+ c_{001} \frac{d}{d\ln R_*} \left(\frac{\s^{2}_{F, s}}{\s_s^2} \right)
\bigg]
\vs
& + A_\alpha \xi^{(1)}_{\nabla^2\alpha L}(r) c_{100} \bigg[
c_{200} \s^{2}_{f, L}  + c_{010} \frac{\s^{2}_{f, s}}{\s^2_{s}}
+ c_{001} \frac{d}{d\ln R_*} \left(\frac{\s^{2}_{f, s} }{\s_s^2} \right)
\bigg] \Bigg\}\,.
\ea
On the other hand, inserting \refeqs{b010}{b001} into \refeq{xihtri} yields
\ba
\xi_h^{\rm tri}(r) = \frac1{\N^2} \Bigg\{& c_{100}^2 \xi_L(r) 
+ 2 c_{100} \left[ c_{200} \s_{F,L}^2 + c_{010} \frac{\s^2_{F,s}}{\s_{s}^2} 
+ c_{001} \frac{d}{d\ln R_*}\left(\frac{\s^2_{F,s}}{\s_s^2} \right)
\right] A_\alpha \xi^{(1)}_{\alpha L}(r) \vs
& + 2 c_{100} \left[c_{200} \s^2_{f,L} + c_{010} \frac{\s_{f,s}^2}{\s_s^2}
+ c_{001} \frac{d}{d\ln R_*}\left(\frac{\s^2_{f,s}}{\s_s^2} \right)
\right] A_\alpha  \frac12 \xi^{(1)}_{\nabla^2\alpha L}(r) \Bigg\}\,.
\ea
We thus find that the trivariate renormalized expression,
\ba
\xi_h^{\rm tri}(r) =\:& b_{100}^2 \xi_L(r) 
+ 2 b_{100} b_{010} A_\alpha \xi^{(1)}_{\alpha L}(r) 
+ 2 b_{100} b_{001} A_\alpha  \frac12 \xi^{(1)}_{\nabla^2\alpha L}(r)
\label{eq:xihtri2}
\ea
has successfully absorbed all $R_L$-dependence from the bare expansion including subleading terms in the squeezed limit (that is, apart from the residual dependence through $\xi_L,\,\xi^{(1)}_{\alpha L},\,\xi^{(1)}_{\nabla^2\alpha L}$, which is negligible on large scales).

\subsection{Total contribution}
\label{sec:sum}

We are now ready to derive the full expression for the two-point tracer
correlation including subleading corrections in the squeezed limit for the
general separable bispectrum \refeq{Bphigen}.  For each $\alpha$, and each of the three cyclic permutations, we define general trivariate bias parameters through
\be
b_{NML}^{(\alpha, ij)} = \frac1{\<n_h\>(0)} \frac{\partial^{N+M+L} \<n_h\>(D,\eps^{(\alpha,ij)},\iota^{(\alpha,ij)})}{\partial D^N\,\partial (\eps^{(\alpha,ij)})^M\, \partial (\iota^{(\alpha,ij)})^L} \Big|_0 \,,
\ee
where $\eps^{(\alpha,ij)},\,\iota^{(\alpha,ij)}$ are scale-dependent rescalings of the density field given by
\ba
\d(\vk_1) \to\:& \d(\vk_1)\left[1 + \eps^{(\alpha, ij)} \frac{F^{(i)}_\alpha(k_1) F^{(j)}_\alpha(k_1)}{P_\phi(k_1)} \right] \vs
\d(\vk_1) \to\:& \d(\vk_1) \left\{ 1 + \iota^{(\alpha, ij)}
\frac1{6 k_1^2 P_m(k_1)}
\left[\tilde F_\alpha^{(i)}(k_1) \left[2 k_1 \tilde F'^{(j)}_\alpha(k_1) + k_1^2 \tilde F''^{(j)}_\alpha(k_1)\right] + (i) \leftrightarrow (j)\right]\right\}\,.
\label{eq:trans_gen}
\ea
Note that $b_{NML}^{(\alpha, ij)} = b_{NML}^{(\alpha, ji)}$.  Then, the renormalized tracer 2-point correlation at tree level is given by
 \ba
\xi_h^{\rm tri}(r) =\:& b_{100}^2 \xi_L(r) 
+ 2 b_{100} \sum_\alpha A_\alpha \left[ 
b_{010}^{(\alpha,23)} \xi^{(1)}_{\alpha L}(r) 
+ \frac12 b_{001}^{(\alpha,23)} \xi^{(1)}_{\nabla^2\alpha L}(r)
+ \{(123) \to (231)\} + \{(123) \to (312) \}\right]\,.
\label{eq:xihtot}
\ea
Note that when going beyond tree level, one in general also has to include mixed contributions simultaneously involving rescalings \refeq{trans_gen} for different $\alpha$ and/or permutations.  \refeq{xihtot} is accurate up to terms of order $(k/k_1)^4$ in the squeezed limit.  We will make this statement more precise in the following sections.  We will discuss the expected relative amplitude of $b_{010}$ and $b_{001}$ below in \refsec{univ}.  Note again that both $b_{010}$ and $b_{001}$ are in general dimensionful [\refeq{dimb01}, \refeq{dimb001}].

\end{widetext}

\subsection{Tracer correlations in Fourier space}
\label{sec:FS}

In Fourier space, the tree-level result \refeq{xihtot} can be phrased in terms of a generalized scale-dependent bias:
\ba
& \frac{P_h(k)}{P_m(k)} = b_{100}^2  + 2 b_{100} \D b(k)
\label{eq:Phtot} \\
& \D b(k) \equiv\: \sum_\alpha A_\alpha \bigg[  b^{(\alpha,23)}_{010} \S_\alpha^{(1)}(k) + \frac12 b^{(\alpha,23)}_{001} k^2 \S_\alpha^{(1)}(k) \vs
& \hspace*{2cm} + \{(123) \to (231)\} + \{(123) \to (312) \}\bigg]
\vs
& \S_\alpha^{(i)}(k) = \frac{F^{(i)}_\alpha(k)}{\M_L(k) P_\phi(k)}\,.
\label{eq:Phtot2}
\ea
Thus for each contribution $\alpha$ in the separable bispectrum \refeq{Bphigen}, 
and for each of the three cyclic permutations, we have in general two independent contributions to the scale-dependent bias.  The leading term in the large-scale limit ($k\to 0$) scales as $\S(k)$, while the subleading term scales as $k^2 \S(k)$.  

\refeq{Phtot} now allows us to estimate the relative magnitude of the subleading term.  For each $\alpha$ and permutation,
\be
\frac{\D b_{\rm subleading}(k)}{\D b_{\rm leading}(k)} = \frac{k^2 b_{001}^{(\alpha,23)}}{2 b_{010}^{(\alpha,23)}}\,.
\label{eq:SLoverL}
\ee
On the other hand, the contributions from different $\alpha$ and permutations $(ijk),\,(lmn)$ scale as
\be
\frac{\D b^{(\beta, l)}(k)}{\D b^{(\alpha, i)}(k)} = \frac{b_{010}^{\beta (mn)}\S^{(l)}_\beta(k)}{b_{010}^{\alpha (jk)} \S^{(i)}_\alpha(k)}\,,
\label{eq:betaoveralpha}
\ee
and similarly for the subleading term.  
Since \refeq{xihtot} and \refeq{Phtot} are derived neglecting terms that, for each contribution $\alpha$ and permutation, are suppressed by $k^4$ relative to the leading contribution, only contributions which are suppressed by less than $k^4$ in \refeq{betaoveralpha} relative to the overall leading contribution $(i,\alpha)$ should be included in \refeq{Phtot}.  

Note further that we have not included the stochasticity in \refeq{Phtot} which in some models of primordial non-Gaussianity can become important on large scales as well \cite{tseliakhovich/etal:10,baumann/etal:12}.

\section{Examples and numerical estimates}
\label{sec:ex}

\subsection{Local non-Gaussianity}

As before, for local non-Gaussianity we consider the two leading cylic permutations as well as the leading contribution from the third permutation.  Thus,
\ba
\xi_h^{\rm tri}(r) =\:& b_{100}^2 \xi_L(r) + 2 b_{100} 2 \fNL \bigg[ 
b^{\rm loc}_{010} \xi_{\phi \d_L}(r) 
\label{eq:xihtotloc}\\
& \hspace*{1cm} + \frac12 b^{\rm loc}_{001} \xi_{\varphi \d_L}(r) 
+ \frac12 b^{(12)\rm loc}_{010} \xi^{(3)}_{\alpha L}(r) \bigg]\,.
\nonumber
\ea
Here, the bias parameters are derived with respect to the rescalings
\ba
b^{\rm loc}_{010}:\quad \d(\vk_1) \to\:& \left[1+\eps \right] \d(\vk_1) \vs
b^{\rm loc}_{001}:\quad\d(\vk_1) \to\:& \left[1+ \eps X(k_1) \right] \d(\vk_1) 
\vs
b^{(12)\rm loc}_{010}:\quad\d(\vk_1) \to\:& \left[1+ \eps P_\phi(k_1) \right] \d(\vk_1)\,,
\label{eq:trans_loc}
\ea
where $X(k_1)$ is defined in \refeq{Xdef}.  Note that the leading term agrees with \cite{PBSpaper}.  In Fourier space, this becomes
\ba
\frac{P_h(k)}{P_m(k)} =\:& b_{100}^2 + 2 b_{100} \D b(k) \vs
\D b(k) =\:& 2\fNL \bigg[  b^{\rm loc}_{010} \M_L^{-1}(k) + \frac12 b^{\rm loc}_{001} k^2 \M_L^{-1}(k) \vs
& \hspace*{1cm} + \frac12 b^{(12)\rm loc}_{010} \M_L^{-1}(k) P_\phi^{-1}(k)\bigg]
\label{eq:Phtotloc}\,.
\ea
Following \refsec{FS}, we can estimate the relative magnitude of the subleading correction and the contribution from the third permutation relative to the leading term as
\ba
\frac{\D b^{\rm loc}_{\rm subleading}(k)}{\D b^{\rm loc}_{\rm leading}(k)} 
=\:& \frac{b_{001}^{\rm loc}}{2 b_{010}^{\rm loc}} k^2
\label{eq:Db12loc}\\
\frac{\D b^{(12)\rm loc}(k)}{\D b^{\rm loc}_{\rm leading}(k)} =\:& 
\frac{b_{010}^{(12)\rm loc}}{b_{010}^{\rm loc}}
\frac{1}{2 P_\phi(k)} 
= \frac{k_0^{-3}\,b_{010}^{(12)\rm loc}}{2 \A_s\,b_{010}^{\rm loc}}
k^3\,,
\nonumber
\ea
where in the last line we have assumed a scale-invariant spectrum of $\phi$ with amplitude $\A_s$ defined at the pivot scale $k_0$,
\be
P_\phi(k) = \A_s \left( \frac{k}{k_0} \right)^{-3}\,.
\ee
Thus, the subleading term is suppressed by a factor of $k^2$ with respect to the leading term, while the contribution from the third permutation is suppressed by $k^3$.  

\subsection{Universal mass function}
\label{sec:univ}

In order to quantitatively assess the importance of the subleading term in \refeq{xihtot} and \refeq{Phtot}, we need to estimate the magnitude of $b_{001}$.  For dark matter halos this can be done accurately through N-body simulations with modified initial conditions following \refeq{trans_gen} [or \refeq{trans_loc} for local non-Gaussianity].  However, a detailed comparison with N-body simulations is beyond the scope of this paper.  Instead, we make use of a generalization of the universal mass function prescription as discussed in Sec.~IV~E of \cite{PBSpaper}.  We write the mean abundance of tracers as 
\ba
\bar n_h =\:& \bar n_h\left(\bar\rho,\sigma_*, J_*\right)
\label{eq:nhsimple} \\
J_* \equiv\:& \frac{d\ln \s_*}{d\ln R_*}\,,
\label{eq:Js}
\ea
where $\s_*$ is the variance of the linear matter density field on scale $R_*$, and $R_*$ is related to the mass $M_*$ through $M_* = 4\pi/3\, \rhob R_*^3$.  The Jacobian $J_*$ is present in order to convert from an interval in $\s_*$ to a mass interval.  In this approximation, $\bar n_h$ is given as a function of the mean density of the Universe and the variance of the density field smoothed on a single scale $R_*$, as well as its derivative with respect to scale.  

\begin{figure}[t!]
\centering
\includegraphics[width=0.49\textwidth]{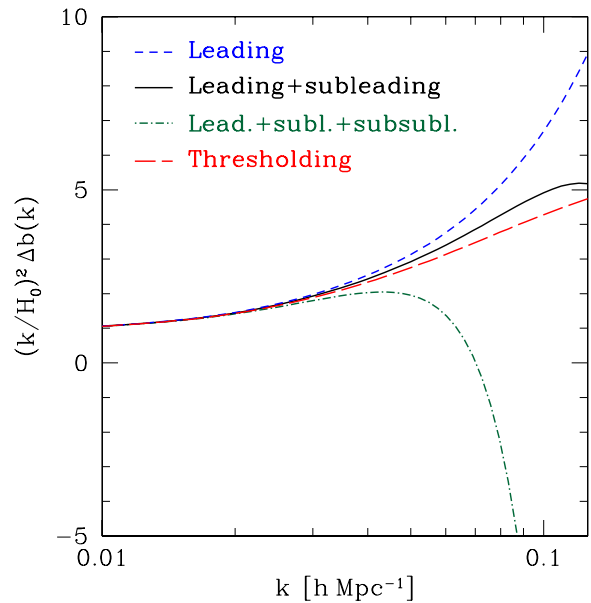}
\caption{Contributions to the scale-dependent bias from local non-Gaussianity ($\fNL=1$) for halos with $M=2\cdot 10^{13}\Msunh$ at $z=0$ ($b_{100}\simeq 0.44$) and assuming a universal mass function from the Sheth-Tormen prescription \cite{sheth/tormen:1999}, scaled by $(k/H_0)^2$ to yield a scale-independent value on large scales.  The dashed line shows the leading term, the solid the leading plus subleading (order $(k/k_*)^2$) term, while the dash-dotted line includes the order $(k/k_*)^3$ term.  The red long-dashed line shows the prediction from the conditional PS mass function (\refsec{thr}).
\label{fig:Db}}
\end{figure}

Let us again consider an individual contribution $\alpha$ and permutation $(123)$.  Under the rescaling \refeq{trans_gen}, $\s_*$ transforms to lowest order as
\ba
\sigma_* \stackrel{\eps}{\longrightarrow}\:& \sigma_*\left[1 + \eps \frac{\sigma^{2(23)}_{\alpha,*}}{\s_*^2}\right] \vs
\sigma_* \stackrel{\iota}{\longrightarrow}\:& 
\sigma_*\left[1 + \iota \frac{\sigma^2_{f,*}}{\s_*^2}\right]\,,
\ea
where $f(k)$ is defined through \refeq{fdef}.  As shown in \cite{PBSpaper} (see also \cite{long}), the scale-dependent biases are then given by
\ba
b_{010} =\:& \left[ \frac1{\bar n_h}\frac{\partial \bar n_h}{\partial\ln \sigma_*}
+ 2\left( \frac{d\ln \s^{2(23)}_{\alpha,*}}{d \ln \s^2_*}
-  1 \right) \right]\frac{\s^{2(23)}_{\alpha,*}}{\s_*^2} \vs
=\:& \left[ b_{010}^{\rm loc} 
+ 2\left( \frac{d\ln \s^{2(23)}_{\alpha,*}}{d \ln \s^2_*}
-  1 \right) 
\right]
\frac{\s^{2(23)}_{\alpha,*}}{\s_*^2} \,,
\label{eq:b010univ}
\ea
and
\ba
b_{001} =\:& \left[ b_{010}^{\rm loc} 
+ 2\left( \frac{d\ln \s^{2}_{f,*}}{d \ln \s^2_*}
-  1 \right) \right]\frac{\s^{2}_{f,*}}{\s_*^2} \,.
\label{eq:b001univ}
\ea
Here, $b_{010}^{\rm loc}$ is the leading scale-dependent bias parameter for local primordial non-Gaussianity, for a tracer following \refeq{nhsimple}.  We have assumed that the tracer density scales linearly with the Jacobian as expected physically.  For such tracers, the leading and subleading bias parameters quantifying the response to general non-local non-Gaussianity are thus directly related to the leading bias parameter for local non-Gaussianity.  If we further specialize to a universal mass function, $\bar n_h = \bar n_h(\rhob, \nu = \d_c/\s_*, J_*)$, then $b_{010}^{\rm loc} = b_{100} \d_c$ (recall that $b_{100}$ is the Lagrangian bias).  

The precise magnitude of the bias coefficients $b_{010},\,b_{001}$ depends on the exact rescaling \refeq{trans_gen}.  Generally, if the main contribution to $\s^{2(23)}_{\alpha,*}$ comes from Fourier modes $k\sim k_*$, then
\be
\s^2_{f,*} \sim k_*^{-2}\,\s^{2(23)}_{\alpha,*}\,.
\ee
Note that depending on the shape of the rescaling $k_*$ does not necessarily have to be of order $1/R_*$.  Thus, for tracers following \refeq{nhsimple}, \refeq{SLoverL} simplifies to
\be
\frac{\D b_{\rm subleading}(k)}{\D b_{\rm leading}(k)} \sim \left(\frac{k}{k_*}\right)^2\,.
\ee
More generally, \refeq{nhsimple} implies that for scale-free bispectra for which $\S^{(i)}_\alpha(k) \propto k^{n^{(i)}_\alpha}$, $k_*$ is the only scale involved, so that other contributions as in \refeq{betaoveralpha} will all be suppressed by powers of $k/k_*$.  Conversely, $k \sim k_*$ indicates the breakdown of the perturbative expansion in the squeezed limit.

\subsection{Numerical results}
\label{sec:num}

\reffig{Db} shows the leading and sub-leading scale-dependent bias contributions for local non-Gaussianity assuming tracers following a universal mass function.  Specifically, we assume a halo mass $M=2\cdot 10^{13}\Msunh$, so that $b_{010}^{\rm loc} = b_{100} \d_c = 0.44$.  We have multiplied the scale-dependent bias by $(k/H_0)^2$ in order to obtain a weakly scale-dependent result.  Note that even the leading term has a residual scale dependence due to the transfer function contained in $\M(k)$.  We also show the contribution $\D b^{(12) \rm loc}(k)$ from the third permutation of the local bispectrum [\refeq{Db12loc}].  The higher order contributions become important as $k \gtrsim 0.05\iMpch$.  Note in particular that the term $\D b^{(12) \rm loc}$ grows rapidly towards smaller scales.  We will discuss this issue in the context of the relation to previous approaches in \refsec{thr}.

\begin{figure}[t!]
\centering
\includegraphics[width=0.49\textwidth]{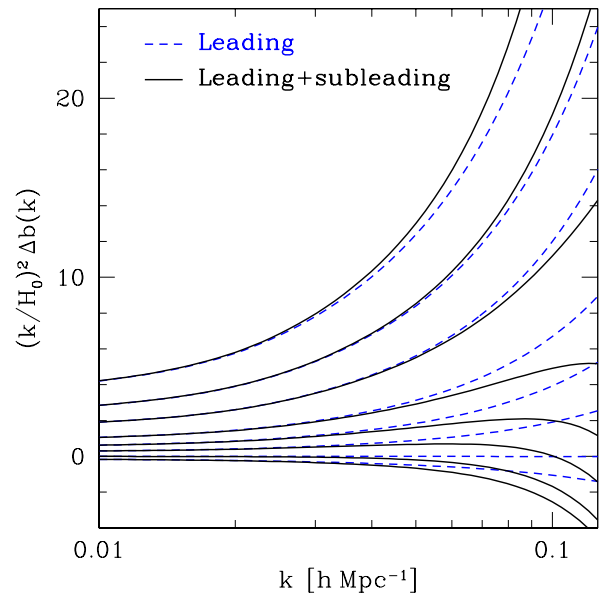}
\caption{Leading and sub-leading contributions to the scale-dependent bias from local non-Gaussianity, as in \reffig{Db}, but for different masses at $z=0$.  The curves shown correspond to, from top to bottom, $M = 2\cdot 10^{14}$, $10^{14}$, $5\cdot 10^{13}$, $2\cdot 10^{13}$, $10^{13}$, $5 \cdot 10^{12}$, $2\cdot 10^{12}$, and $10^{12}\Msunh$, respectively.
\label{fig:Db_M}}
\end{figure}

\reffig{Db_M} shows the leading and leading+subleading contributions to the scale-dependent bias for a range of masses from $10^{12}$ to $2\cdot 10^{14}\Msunh$.  Clearly, the typical scale at which the subleading correction becomes important does not depend sensitively on the mass.  Generally, the correction is more important at lower masses, specifically around $M_*$ where the Lagrangian bias $b_{100}$ vanishes.  For vanishing $b_{100}$, the leading term vanishes, whereas the subleading term does not disappear entirely due to the non-vanishing derivative of $\ln \s_X^2$ with respect to $\ln\s_*^2$.  Since $\s_{X,*}^2$ scales more weakly with $R_*$ than $\s_{*}^2$, this derivative is less than one leading to a suppression of the scale-dependent bias when including the subleading correction.

In principle, a general tracer could lead to very different numerical results, i.e. much larger or smaller subleading corrections.  However, given the accuracy of universal mass functions of $10-20$\% at least for dark matter halos, we expect the magnitude of the corrections as shown in \reffig{Db_M} to be typical.

\section{Relation to previous results}
\label{sec:thr}

We now make the connection to previous results on the scale-dependent bias for general non-Gaussianity beyond the squeezed limit.  
In \citet{long}, the scale-dependent bias was derived by applying a conditional mass function approach to the Press-Schechter (PS) mass function \cite{press/schechter:1974}.  The non-Gaussianity was taken into account by applying an Edgeworth expansion to the Gaussian PDF of the density field.  As shown in \cite{long}, the scale-dependent bias defined through \refeq{Phtot} is in this case given by
\ba
\D b_{\rm PS}(k) =\:& 
 \left[b_1 \d_c + 2\frac{\partial\ln \F_*^{(3)}(k)}{\partial\ln \s_*^2} \right] 2\F_*^{(3)}(k) \M_L^{-1}(k)
\,.
\label{eq:Db_thr}
\ea
Here, a subscript $*$ denotes filtering with a tophat of radius $R_*$, the Lagrangian radius corresponding to the mass scale of the tracer as in \refsec{univ} (this was denoted as $R_s$ in \cite{long}).  
Note the derivative with respect to $\ln\s_*^2$ rather than $\ln\s_*$ as written in \cite{long}.  
Here, the function $\F_*^{(3)}(k)$ is given by an integral over the bispectrum,
\ba
\F_*^{(3)}(k) = \frac1{4\s_*^2 P_\phi(k)} \int \frac{d^3 k_1}{(2\pi)^3} &
\M_*(k_1) \M_*(|\vk+\vk_1|) \vs
& \times B_\phi(\vk,\vk_1,-\vk-\vk_1)\,,
\nonumber
\ea
where $\M_*(k) = \M(k) \tilde W_*k)$.  The function $\F_*^{(3)}$ can then be directly related to the squeezed three-point function 
\ba
& \< \d_L(1) \d_*^2(2) \> = 4 \s_*^2 \int \frac{d^3 k}{(2\pi)^3} \M_L(k) P_\phi(k) \F_*^{(3)}(k) e^{i\vk \vr} \vs
& = \sum_\alpha A_\alpha \bigg\{2 \xi^{(1)}_{\alpha L}(r) 
\s^{2(23)}_{\alpha *}
+  \xi^{(1)}_{\nabla^2\alpha L}(r) \s^{2(23)}_{X \alpha *} + 2\:{\rm perm.}
\bigg\}\,,
\nonumber
\ea
again up to order $(k/k_*)^4$, see for example the derivation leading up to \refeq{diffterms_genA}.  In Fourier space, this relation becomes
\begin{widetext}
\ba
2\F_*^{(3)}(k) =\:& \frac1{P_\phi(k)} \sum_\alpha A_\alpha \bigg\{
 F^{(1)}_\alpha(k) \frac{\s^{2(23)}_{\alpha *}}{\s_*^2}
+ \frac12 k^2 F^{(1)}_\alpha(k) 
\frac{\s^{2(23)}_{X \alpha *}}{\s_*^2} + \{ (123) \to (231) \} + \{ (123) \to (312) \}
\bigg\}\,.
\ea
\refeq{Db_thr} then becomes
\ba
\D b_{\rm PS}(k) =\:&  \frac1{P_\phi(k) \M_L(k)} \sum_\alpha A_\alpha \bigg\{
 F^{(1)}_\alpha(k) \left[b_1 \d_c + 2\frac{\partial}{\partial\ln \s_*^2} \right] 
\frac{\s^{2(23)}_{\alpha *}}{\s_*^2}
+ \frac12 k^2 F^{(1)}_\alpha(k) 
\left[b_1 \d_c + 2\frac{\partial}{\partial\ln \s_*^2} \right] 
\frac{\s^{2(23)}_{X \alpha *}}{\s_*^2} + 2\:{\rm perm.}
\bigg\} \vs
=\:& \sum_\alpha A_\alpha \bigg\{
 \S^{(1)}_\alpha(k) \left[b_1 \d_c + 2\frac{\partial}{\partial\ln \s_*^2} \right] 
\frac{\s^{2(23)}_{\alpha *}}{\s_*^2}
+ \frac12 k^2 \S^{(1)}_\alpha(k) 
\left[b_1 \d_c + 2 \frac{\partial}{\partial\ln \s_*^2} \right] 
\frac{\s^{2(23)}_{X \alpha *}}{\s_*^2} + 2\:{\rm perm.}
\bigg\}\,,
\ea
\end{widetext}
using the definitions after \refeq{Phtot}.  Comparing with \refeq{Phtot}, we can now read off the bias parameters
\ba
b_{010}^{(\alpha,23)} =\:& \left[b_1 \d_c + 2\frac{\partial}{\partial\ln \s_*^2} \right] 
\frac{\s^{2(23)}_{\alpha *}}{\s_*^2} \vs
=\:& \left[b_1 \d_c + 2\left(\frac{\partial\ln \s^{2(23)}_{\alpha *}}{\partial\ln \s_*^2} -1 \right)\right] 
\frac{\s^{2(23)}_{\alpha *}}{\s_*^2}\,,
\ea
\ba
b_{001}^{(\alpha,23)} =\:& \left[b_1 \d_c + 2\left(\frac{\partial\ln \s^{2(23)}_{X \alpha *}}{\partial\ln \s_*^2} -1 \right)\right] 
\frac{\s^{2(23)}_{X \alpha *}}{\s_*^2} \,.
\ea
We see that both the leading and subleading bias parameters derived from the conditional PS mass function agree with those expected from a general universal mass function [\refeqs{b010univ}{b001univ}, with $b_{010}^{\rm loc} = b_{100} \d_c$].  Fundamentally, this is a consequence of the fact that in the Press-Schechter approach, as in general for universal mass functions, there is only one scale $R_*$ that enters the description of tracer statistics.  We thus expect this result to hold at higher order in $k/k_*$ as well.

\begin{figure}[t!]
\centering
\includegraphics[width=0.49\textwidth]{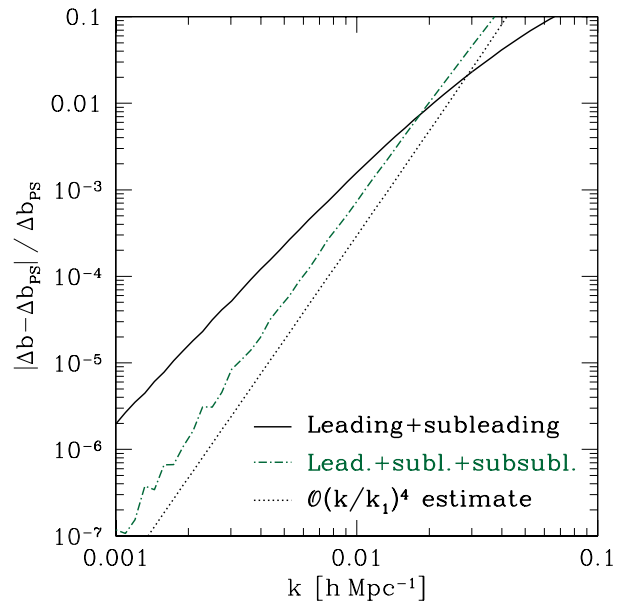}
\caption{Fractional difference between the contributions in \refeq{Phtotloc}, evaluated for a universal mass function, and the result for the conditional PS mass function \refeq{Db_thr} from \cite{long} for local primordial non-Gaussianity.  The black line solid line shows the residuals when including the leading and subleading (order $(k/k_*)^2$) contributions, while the green dash-dotted line also includes the order $(k/k_*)^3$ contribution from the last line of \refeq{Phtotloc}.  The dotted line shows a rough estimate of the order $(k/k_*)^4$ correction (see text).  We have again assumed $M = 2\cdot 10^{13}\Msunh$ and $z=0$, although the results are essentially independent of the mass.
\label{fig:Db_log}}
\end{figure}

Since \refeq{Db_thr} does not involve a perturbative expansion in the ratio of wavenumbers $k/k_*$, it can serve as a useful guide as to where this expansion breaks down.  \reffig{Db_log} shows the residuals when including the leading and subleading terms from \refeq{Phtotloc}.   We see a residual which scales as $k^3$ for small $k$.  When including the term from the last line of \refeq{Phtotloc}, we see that the residuals become even smaller and scale as $k^4$ as expected.  Some numerical artefacts are visible when the residuals become of order $10^{-6}$ or smaller.  These are due to the numerical derivative performed when evaluating \refeq{Db_thr}.  We can in fact perform a rough estimate of the expected correction order $(k/k_*)^4$ in \refeq{Phtotloc}, via
\ba
\D b_{\rm NNLO\:est.} =\:& \fNL b_{100} \d_c \frac{\s_{-4,*}^2}{\s_*^2} k^4 \M_L^{-1}(k) \label{eq:NNLOest}\\
\s_{-4,*}^2 =\:& \frac1{2\pi^2} \int_{k_{\rm min}}^\infty k^2 dk\:k^{-4} P_m(k) \tilde W_L^2(k)\,.\nonumber
\ea
Note that due to the logarithmic divergence we need to introduce a low-$k$ cutoff in $\s_{-4,*}^2$.  This is likely to be an artefact of the universal mass function prescription, where the leading effect of a change in the small-scale power spectrum shape on the tracer density is given by this formally divergent spectral moment.  In reality, tracers will have a finite response to such a change.  Here we choose $k_{\rm min} = 0.01\iMpch$, corresponding to the turnover in $P_m(k)$.  \refeq{NNLOest} is in any case only to be seen as a very rough estimate.  This contribution is shown as dotted line in \reffig{Db_log}, making clear that the residual, after taking into account all terms in \refeq{Phtotloc}, indeed scales as $k^4$ on large scales.  

\reffig{Db_log} shows that for the local model, the order $(k/k_*)^3$ correction becomes comparable to the lower order corrections at $k\sim 0.02\iMpch$, signaling a breakdown of the perturbative expansion there, even though the fractional deviation from the full result \refeq{Db_thr} when including terms up to order $(k/k_*)^2$ remain at 10\% or less all the way to $k\sim 0.1\iMpch$.  

In summary, the conditional PS mass function results derived in \cite{long} are consistent for tracers following a universal mass function, in the sense that they match the results from the general renormalization approach when restricted to universal mass functions.  Note that this holds once the polynomials in $\d_c/\s_*$ are replaced with bias parameters, as described in \cite{long}.  In this context, the key advantage of the PS approach is that it sums over all powers of $k/k_*$, without relying on a perturbative expansion in this parameter.

On the other hand, realistic tracers will not simply depend on the variance of the density field on a single scale, thus breaking the relation between the bias parameters $b_{001},\,b_{010}$, and $b_{001}$.  Furthermore, there are other scale-dependent biases which contribute at the same order as the subleading correction $b_{001}$.  We will turn to this issue next.

\section{Scale-dependent bias beyond the large-scale limit}
\label{sec:other}

We have seen that beyond the squeezed limit, there is a subleading correction to the scale-dependent bias from primordial non-Gaussianity that scales as $k^2$ relative to the leading term.  In addition to this correction however, we expect two additional contributions that are leading order in $\fNL$ (i.e., in the primordial bispectrum), and have the same scaling with $k$.

First, as shown in \cite{mcdonald/roy:2009,PBSpaper}, non-locality in the formation of tracers generically induces a dependence on the curvature of the density field, leading to a contribution of
\be
\xi_h(r) \supset b_{\nabla^2\d} b_{010} \< \nabla^2 \d_L(1) y_*(2)\>\,.
\label{eq:xinabd}
\ee
This contribution will also serve to absorb the residual $R_L$-dependence present in $\xi^{(i)}_{\alpha L}(r),\,\xi^{(i)}_{\nabla^2\alpha L}(r)$ in a similar way as discussed in \cite{PBSpaper}.  In Fourier space, \refeq{xinabd} corresponds to a contribution to the scale-dependent bias of the form
\be
\frac{P_h(k)}{P_m(k)} \supset 2 b_{\nabla^2\d} k^2 \D b_{\rm lead}(k)\,,
\ee
where $\D b_{\rm lead}$ is the leading contribution to the non-Gaussian scale-dependent bias from \refeq{Phtot2}.  If $L_\d$ is the scale of non-locality of the tracer (in terms of its dependence on the matter density), then $b_{\nabla^2\d} \sim L_\d^2$, so that this additional contribution scales as $(k L_\d)^2$ relative to the leading term.

Throughout the discussion of the non-Gaussian case in \cite{PBSpaper} and here, we have assumed that the tracer density is a purely local function of $y_*$, the parameter which quantifies the amplitude of small-scale fluctuations (in this paper, we have also introduced a local dependence on $\mu_* = dy_*/d\ln R_*$).  In general, however, one also expects that tracers depend on the amplitude of small-scale fluctuations in some finite region of size $L_y$.  Then, in straightforward analogy with the density case, we also need to allow for a bias $b_{\nabla^2 y}$ with respect to $\nabla^2 y_*$, where $b_{\nabla^2 y} \sim L_y^2$.  That is, strictly speaking we need to generalize each of the rescalings in \refeq{trans_gen} to be spatially dependent, $\eps \to \eps\, \vx^2$.  Schematically, this leads to a contribution to the tracer correlation of
\be
\xi_h(r) \supset b_{100} b_{\nabla^2 y} \< \d_L(1) \nabla^2 y_*(2)\>\,,
\ee
which, in Fourier space, becomes
\be
\frac{P_h(k)}{P_m(k)} \supset 2 b_{100} b_{\nabla^2 y} k^2 \frac{\D b_{\rm lead}(k)}{b_{010}}\,.
\ee
This contribution thus scales as $(k L_y)^2$ relative to the leading term.  

Typically, one might expect $L_\d \sim L_y \sim R_*$, where $R_*$ is the Lagrangian radius of the region that collapses to form the tracer.  On the other hand, the length scale we found for the subleading terms in the squeezed limit is $1/k_* \sim 50\Mpch$, suggesting that this correction is somewhat more important than the other contributions described in this section for typical tracers for which $R_* = 1-10\Mpch$.  However, the value of $k_*$ depends on the specific type of non-Gaussianity considered, and in general all three subleading contributions to the scale-dependent halo bias can be comparable in magnitude.  Thus, 
if one of them is included (even implicitly as for example in the conditional PS mass function result), then all of them should be included for consistency, unless one can show that the subleading contribution dominates over the other contributions described in this section.

\section{Conclusions}
\label{sec:concl}

We have derived the subleading contributions to the scale-dependent bias $\D b(k)$ 
of large-scale structure tracers for a general separable primordial bispectrum.  The leading contribution is given by the scaling of the bispectrum in the squeezed limit, $\lim_{k\to 0} B_\phi(k,k_s,|\vk_s+\vk|)$, and the scale-dependence of the subleading contribution is suppressed by a factor of $k^2$ relative to this term.  This subleading contribution is important to quantify, since it tells us at which $k$ the usual squeezed-limit result ceases to be accurate.  For local non-Gaussianity and tracers following a universal mass function, we found that this happens at $k \sim 0.02\iMpch$, although the first two leading terms provide an excellent approximation to the result from the conditional PS mass function up to $k \lesssim 0.1\iMpch$.  Our approach is independent of any assumptions on the tracers apart from a finite scale of non-locality.

Throughout, our results have been phrased in terms of the three-point function of the primordial perturbations, which is the standard result of computations performed for particular inflationary models, thus allowing for a direct application of the results of this paper to models of inflation.  In contrast, several previous papers \cite{giannantonio/porciani:2010,schmidt/kamionkowski:2010,scoccimarro/etal:2012} have employed a fictitious Gaussian field mapped to the physical field via a quadratic kernel.  In this latter approach, the effect on large-scale structure tracers is mediated by modulated spectral moments of the density field, e.g. $\s_*^2|_\phi$.  This approach is complicated by the fact that the kernel is not uniquely determined by the bispectrum, so that additional constraints need to be imposed \cite{scoccimarro/etal:2012}.  However, the derivation in this paper can also be applied to the kernel approach in a straightforward way.  Note that while the bispectrum specifies the kernel uniquely in the squeezed limit \cite{schmidt/kamionkowski:2010} (for non-divergent kernels), this is no longer the case when including subleading terms.  Thus, different kernels which yield the same bispectrum are expected to make different predictions for the subleading contribution to the scale-dependent bias.  We leave this issue for future work.

In addition to the subleading contribution in the squeezed limit, we have pointed out two other contributions that will contribute at linear order in $\fNL$ and with the same tree-level $k$-dependence (\refsec{other}).  These contributions should be included (or at least carefully considered) when putting constraints on non-Gaussianity using large-scale structure statistics on intermediate and small scales, i.e. for $k \gtrsim 0.05 \iMpch$.

It is straightforward to extend the squeezed limit expansion presented here to higher order in $k$.  In that case, one needs to add a dependence of the tracer density on another property of the density field further quantifying the sensitivity to the amplitude of small-scale fluctuations as a function of scale.  One possible choice would be $d^2 y_*/ d(\ln R_*)^2$.  

Another straightforward extension is the inclusion of higher primordial $N$-point functions.  For example, in the presence of a primordial four-point function both the linear and quadratic bias become scale-dependent \cite{long,smith/etal:12}, and one needs to take into account the dependence of the tracer density on the skewness of the density field.  No conceptually new issue arises, and the calculation will closely follow the one presented here.

Finally, we have shown how the coefficient of both leading and subleading terms in the scale-dependent bias for a general bispectrum can be derived for tracers identified in N-body simulations, by running simulations with modified initial conditions [\refeq{trans_gen}].  This will allow for a precise test of the accuracy of the universal mass function prediction for dark matter halos in the context of non-Gaussian halo bias.  We leave this for future work.

\acknowledgments

The author thanks Kendrick Smith and Svetlin Tassev for helpful discussions.  This work was supported by NASA through Einstein Postdoctoral Fellowship grant number PF2-130100 awarded by the Chandra X-ray Center, which is operated by the Smithsonian Astrophysical Observatory for NASA under contract NAS8-03060.

\begin{widetext}
\appendix

\section{Derivation of \refeq{diffterms_gen} and \refeq{dy_gen}}
\label{app:corrderivN}

In this appendix we derive the next-to-leading order squeezed-limit expressions \refeq{diffterms_gen} and \refeq{dy_gen}.  As described in \refsec{new}, we assume that the bispectrum $B_\phi$ is given in separable form,
\ba
B_\phi(k_1, k_2, k_3) = \sum_\alpha \bigg[ & F^{(1)}_\alpha(k_1) F^{(2)}_\alpha(k_2) F^{(3)}_\alpha(k_3) + F^{(1)}_\alpha(k_1) F^{(3)}_\alpha(k_3) F^{(2)}_\alpha(k_2) \vs
& + F^{(2)}_\alpha(k_1) F^{(3)}_\alpha(k_2) F^{(1)}_\alpha(k_3) + F^{(2)}_\alpha(k_1) F^{(1)}_\alpha(k_3) F^{(3)}_\alpha(k_2) \vs
& + F^{(3)}_\alpha(k_1) F^{(1)}_\alpha(k_2) F^{(2)}_\alpha(k_3) + F^{(3)}_\alpha(k_1) F^{(2)}_\alpha(k_3) F^{(1)}_\alpha(k_2) 
\bigg]\,,
\ea
where the 6 permutations guarantee the symmetry of $B_\phi$ in its arguments. This leads to \refeq{diffterms2}.  We now expand the $k_1$ integrand in powers of $k/k_1$ up to second order.  For notational simplicity,
we only consider the contribution from a single term $\alpha$ and two permutations
$(2) \leftrightarrow (3)$, and drop the subscript $\alpha$ for the moment.  This yields
\ba
& \int \frac{d^3 k_1}{(2\pi)^3} \bigg[\tilde F^{(2)}(\vk_1 ) \tilde F^{(3)}(-\vk_1 - \vk) + (2) \leftrightarrow (3) \bigg] = \vs
&\int \frac{d^3 k_1}{(2\pi)^3} \bigg[\tilde F^{(2)}(\vk_1) 
\left(1 - k^i \partial_i + \frac12 k^i k^j \partial_i \partial_j \right)
\tilde F^{(3)}(-\vk_1) + (2) \leftrightarrow (3) \bigg] \vs
& = \int \frac{d^3 k_1}{(2\pi)^3} \bigg[\tilde F^{(2)}(\vk_1)\tilde F^{(3)}(-\vk_1) 
- \tilde F^{(3)}(-\vk_1)  k^i \partial_i \tilde F^{(2)}(\vk_1) 
+ \frac12 \tilde F^{(3)}(-\vk_1)  k^i k^j \partial_i \partial_j \tilde F^{(2)}(\vk_1) + (2) \leftrightarrow (3) \bigg]\,,
\ea
where all derivatives are with respect to $\vk_1$.  We now perform an integration
by parts for the second and last terms.  The former term (linear in $\vk$) cancels
with its permutation $(2)\leftrightarrow (3)$, yielding
\ba
\int \frac{d^3 k_1}{(2\pi)^3} \bigg[2 \tilde F^{(2)}(\vk_1)\tilde F^{(3)}(-\vk_1) 
+ \frac12  \tilde F^{(3)}(-\vk_1)  k^i k^j \partial_i \partial_j \tilde F^{(2)}(\vk_1)
+ \frac12  \tilde F^{(2)}(\vk_1)  k^i k^j \partial_i \partial_j \tilde F^{(3)}(-\vk_1) \bigg]\,.
\ea
Note that we have not used that $\tilde F^{(i)}(\vk_1) = \tilde F^{(i)}(k_1)$ so far.  We now use this fact however to obtain, defining $\mu = \hat\vk\cdot\hat\vk_1$,
\be
k^i k^j \partial_i \partial_j F(k_1) = \frac{k^2}{k_1^2} \left[(1-\mu^2) k_1 F'(k_1) + \mu^2 k_1^2 F''(k_1)\right]\,,
\ee
where we have denoted derivatives with respect to $k_1$ with primes.  
We thus obtain, up to terms of order $k^4/k_1^4$ (cubic terms drop out just like the
linear terms did)
\ba
 \<\d_L(1)\d_L^2(2)\> = \sum_\alpha \bigg\{& \int \frac{d^3 k}{(2\pi)^3} \tilde F^{(1)}_\alpha(k) e^{i\vk\cdot\vr}
\int \frac{d^3 k_1}{(2\pi)^3} 
\bigg[2 \tilde F_\alpha^{(2)}(k_1)\tilde F_\alpha^{(3)}(k_1) \vs
& \hspace*{3.5cm} + \frac12  \frac{k^2}{k_1^2} 
\tilde F_\alpha^{(2)}(k_1)\left[(1-\mu^2) k_1 \tilde F'^{(3)}_\alpha(k_1) + \mu^2 k_1^2 \tilde F''^{(3)}_\alpha(k_1)\right]
\vs
& \hspace*{3.5cm} + \frac12  \frac{k^2}{k_1^2} 
\tilde F_\alpha^{(3)}(k_1)\left[(1-\mu^2) k_1 \tilde F'^{(2)}_\alpha(k_1) + \mu^2 k_1^2 \tilde F''^{(2)}_\alpha(k_1)\right]
\bigg] \vs
& + \{ (123) \to (231) \} + \{ (123) \to (312) \}
\bigg\}\,.
\ea
This expression is now in the desired separable form, and the $\mu$ integral becomes trivial.  We now introduce some notation,
\ba
\xi^{(i)}_{\alpha L}(r) \equiv\:& \int \frac{d^3 k}{(2\pi)^3} \M_L(k) F^{(i)}_\alpha(k) e^{i\vk\cdot\vr} \vs
\xi^{(i)}_{\nabla^2\alpha L}(r) \equiv\:& \int \frac{d^3 k}{(2\pi)^3} \M_L(k) k^2\,F^{(i)}_\alpha(k) e^{i\vk\cdot\vr} \vs
\s^{2(ij)}_{\alpha L} \equiv\:& \frac1{2\pi^2} \int_0^\infty dk_1\,k_1^2 \M_L^2(k_1)
F_\alpha^{(i)}(k_1) F_\alpha^{(j)}(k_1) \vs
\s^{2(ij)}_{X\alpha L} \equiv\:& \frac1{2\pi^2} \int_0^\infty dk_1\,
\frac12\left\{\tilde F_\alpha^{(i)}(k_1)\left[\frac23 k_1 \tilde F'^{(j)}_\alpha(k_1) + \frac13 k_1^2 \tilde F''^{(j)}_\alpha(k_1)\right] + (i) \leftrightarrow (j) \right\} \vs
=\:& \frac1{12\pi^2} \int_0^\infty dk_1\,
\left\{\tilde F_\alpha^{(i)}(k_1)\left[2 k_1 \tilde F'^{(j)}_\alpha(k_1) + k_1^2 \tilde F''^{(j)}_\alpha(k_1)\right] + (i) \leftrightarrow (j) \right\}\,.
\ea
Note that for all spectral moments, $\s^{2(ij)} = \s^{2(ji)}$.  
This allows us to write \refeq{diffterms_gen} in the compact form
\ba
 \<\d_L(1)\d_L^2(2)\> =\:& \sum_\alpha \bigg\{2\xi^{(1)}_{\alpha L}(r) \s^{2(23)}_{\alpha L}
+ \xi^{(1)}_{\nabla^2\alpha L}(r) \s^{2(23)}_{X \alpha L} + 2\:{\rm perm.}
\bigg\}\,,
\label{eq:diffterms_genA}
\ea
where the two permutations stand for the cyclic permutations $123 \to 231,\:312$.  

We now turn to \refeq{dy_gen}.  This correlator is given by
\ba
\<\d_L(1) y_*(2)\> =\:& \frac1{2\s_s^2} \< \d_L(1) \d_s^2(2) \> 
\vs
=\:& \frac1{2\s_s^2} \int \frac{d^3 k}{(2\pi)^3} \M_L(k) e^{i\vk\cdot\vr}
\int \frac{d^3 k_1}{(2\pi)^3} \M_s(|\vk_1|) \M_s(|\vk_1+\vk|) 
B_\phi(|\vk|,|\vk_1|,|\vk_1+\vk|)\,,
\label{eq:difftermsB}
\ea
where $\M_s(k) = \tilde W_s(k) \M(k)$.  This expression is very similar to
\refeq{diffterms2}, the only difference being that the filter functions
under the $k_1$ integral involve $\tilde W_s$ rather than $\tilde W_L$.  
\refeq{diffterms_genA} then straightforwardly translates to
\ba
 \<\d_L(1) y_*(2)\> =\:& \sum_\alpha A_\alpha \bigg\{\xi^{(1)}_{\alpha L}(r) 
\frac{\s^{2(23)}_{\alpha s}}{\s_s^2}
+ \frac12 \xi^{(1)}_{\nabla^2\alpha L}(r) \frac{\s^{2(23)}_{X \alpha s}}{\s_s^2} + 2\:{\rm perm.}
\bigg\}\,,
\label{eq:dy_genA}
\ea
where
\ba
\s^{2(ij)}_{\alpha s} \equiv\:& \frac1{2\pi^2} \int_0^\infty dk_1\,k_1^2 \M_s^2(k_1)
F_\alpha^{(i)}(k_1) F_\alpha^{(j)}(k_1) \vs
\s^{2(ij)}_{X\alpha s} \equiv\:& \frac1{12\pi^2} \int_0^\infty dk_1\,
\left\{\M_s(k_1)  F_\alpha^{(i)}(k_1)\left[2 k_1 \left(\M_s F^{(j)}_\alpha\right)'_{k_1} + k_1^2 \left(\M_s F^{(j)}_\alpha\right)''_{k_1}\right] + (i) \leftrightarrow (j) \right\}\,.
\ea

\end{widetext}

\bibliography{PBSng}

\end{document}